\documentclass[prl,aps,superscriptaddress,twocolumn,showpacs,nofootinbib,floatfix,preprintnumbers]{revtex4}

\usepackage{graphicx}

\begin{document}

\preprint{
  KEK-CP-140
  }

\title{
  $B^0-\bar{B}^0$ mixing in unquenched lattice QCD
  }

\newcommand{\Tsukuba}{
  Institute of Physics, 
  University of Tsukuba, 
  Tsukuba 305-8571, 
  Japan}

\newcommand{\RCCP}{
  Center for Computational Physics, 
  University of Tsukuba, 
  Tsukuba 305-8577, 
  Japan}

\newcommand{\ICRR}{
  Institute for Cosmic Ray Research, 
  University of Tokyo, 
  Kashiwa 277-8582, 
  Japan}

\newcommand{\KEK}{
  High Energy Accelerator Research Organization (KEK), 
  Tsukuba 305-0801, 
  Japan}

\newcommand{\Hiroshima}{
  Department of Physics, 
  Hiroshima University,
  Higashi-Hiroshima 739-8526, Japan}

\newcommand{\YITP}{
  Yukawa Institute for Theoretical Physics, 
  Kyoto University, 
  Kyoto 606-8502, Japan}

\newcommand{\RBRC}{
  RIKEN BNL Research Center,
  Brookhaven National Laboratory,
  Upton, NY 11973, U.S.A.}

\author{S.~Aoki}
\affiliation{\Tsukuba}

\author{M.~Fukugita}
\affiliation{\ICRR}

\author{S.~Hashimoto}
\affiliation{\KEK}

\author{K-I.~Ishikawa}
\affiliation{\Tsukuba}
\affiliation{\RCCP}

\author{N.~Ishizuka}
\affiliation{\Tsukuba}
\affiliation{\RCCP}

\author{Y.~Iwasaki}
\affiliation{\Tsukuba}

\author{K.~Kanaya}
\affiliation{\Tsukuba}

\author{T.~Kaneko}
\affiliation{\KEK}

\author{Y.~Kuramashi}
\affiliation{\KEK}

\author{M.~Okawa}
\affiliation{\Hiroshima}

\author{T.~Onogi}
\affiliation{\YITP}

\author{N.~Tsutsui}
\affiliation{\KEK}

\author{A.~Ukawa}
\affiliation{\Tsukuba}
\affiliation{\RCCP}

\author{N.~Yamada}
\affiliation{\RBRC}

\author{T.~Yoshi\'{e}}
\affiliation{\Tsukuba}
\affiliation{\RCCP}

\collaboration{JLQCD Collaboration}
\noaffiliation

\date{\today}

\begin{abstract}
  We present an unquenched lattice calculation for the
  $B^0-\bar{B}^0$ transition amplitude.  The calculation, 
  carried out at an inverse lattice spacing $1/a$ = 2.22(4)~GeV, 
  incorporates two flavors of dynamical quarks described by the
  fully $O(a)$-improved Wilson fermion action   
  and heavy quarks described by NRQCD.
  A particular attention is paid to the uncertainty that
  arises from the chiral extrapolation, especially the effect of pion loops, 
  for light quarks, which we find could be sizable for the leptonic
  decay of the $B_d$ meson, whereas
  it is small for the $B_s$ meson and the $B$ parameters. 
  We obtain 
  $f_{B_d}=191(10)(^{+12}_{-22})$ MeV,
  $f_{B_s}/f_{B_d}=1.13(3)(^{+13}_{-\ 2})$,
  $B_{B_d}(m_b)=0.836(27)(^{+56}_{-62})$,
  $B_{B_s}/B_{B_d}=1.017(16)(^{+56}_{-17})$,
  $\xi=1.14(3)(^{+13}_{-\ 2})$,
  where the first error is statistical, and the second is
  systematic, including 
  uncertainties due to chiral extrapolation, 
  finite lattice spacing, 
  heavy quark expansion and perturbative operator matching.
\end{abstract}

\maketitle

The unitarity test for the Cabibbo-Kobayashi-Maskawa (CKM)
matrix entered a new era with the BaBar and Belle
measurements of the angle $\phi_1$ \cite{Aubert:2002ic,Abe:2002px}.
The test requires the
determination of the other angles and the sides of the
unitarity triangle, the precision of the latter being
limited by uncertainties in hadronic matrix elements.
Lattice QCD in principle offers a model-independent
calculation of such matrix elements.
Those matrix elements provided by lattice calculations to date,
however, are based on the quenched approximation,
blindly hoping that quenching does not introduce 
large errors. 

Simulations including creation and annihilation of
a quark anti-quark pair in the vacuum have become feasible 
only recently. 
In this letter we present an unquenched lattice calculation
of the hadronic matrix elements appearing in the
$B^0-\bar{B}^0$ mixing amplitude, which is needed in the 
determination of the CKM matrix element $|V_{td}|$ from the
mass difference $\Delta M_d$
\cite{Hagiwara:fs}.
The matrix element is parametrized as
$\langle\bar{B}_q^0|
\bar{b}\gamma_\mu(1-\gamma_5)q
\bar{b}\gamma_\mu(1-\gamma_5)q
|B^0_q\rangle \equiv
\frac{8}{3}f_{B_q}^2 B_{B_q} M_{B_q}^2$,
where $f_{B_q}$ is the $B$ meson decay constant and
$B_{B_q}$ denotes the $B$ parameter 
($q$ represents $d$ or $s$ quark).
Our prime interest is to calculate $B_{B_q}$, as 
unquenched calculations of $f_{B_q}$ are already 
available \cite{Yamada:2002wh}. We include, however,
the calculation of $f_{B_q}$ to provide a consistent set
of the matrix element for $B^0-\bar{B}^0$ mixing.

With current lattice calculations, 
systematic uncertainties due to the discretization error
and the perturbative matching between continuum and lattice
operators amount to 10--20\%. 
One may improve the accuracy of $|V_{td}|$ by studying
the ratio $\Delta M_s/\Delta M_d$ if 
$B_s-\bar{B}_s$ mixing is measured.
The relevant quantity is 
$\xi\equiv(f_{B_s}\sqrt{B_{B_s}})/(f_{B_d}\sqrt{B_{B_d}})$, 
where many of the systematic uncertainties cancel in the
lattice calculation.

A remaining major uncertainty arises from the
chiral extrapolation of the lattice simulation which  
is made with relatively heavy dynamical quarks.  
One may resort to chiral perturbation theory (ChPT)
as a theoretical guide for the extrapolation.
The problem is that the currently available 
lattice data do not show the logarithmic behavior
expected from long-distance pion loops in ChPT
\cite{Hashimoto:2002vi}.
It is in the scope of the present work to discuss
the uncertainty in the the matrix elements 
associated from the chiral extrapolation
in the absence of the observable logarithmic
behavior. 

The calculation is carried out on the unquenched gauge
configurations generated at $\beta$ = 5.2 on a 
$20^3\times 48$ lattice.
Two flavors of dynamical quarks for the $u$ and $d$ quarks
are simulated at five values of quark mass in the range 
$(0.7 -2.9) m_s$ with
$m_s$ the physical strange quark mass.
This corresponds to the pseudoscalar to vector mass
ratio of 0.6--0.8.
The hopping parameter chosen is  
$K_{sea}$ = 0.1340, 0.1343, 0.1346, 0.1350, and 0.1355. 
For each sea quark mass, 1,200 configurations are
accumulated for measurements from 12,000 HMC trajectories
separated by 10 trajectories.
The lattice spacing we adopt is determined from $\rho$ meson
mass and equals 2.22(4)~GeV after the extrapolation to the
chiral limit. 
This value is consistent with   
2.19($^{+7}_{-5}$)~GeV from the Sommer scale $r_0$ assuming 
the physical value of 0.49~fm, and 2.25(5)~GeV from $f_K$ 
(with an additional $O(5\%)$ error 
from the perturbative matching).  
This suggests that the large width of $\rho$ may not
seriously affect the chiral extrapolation of the $\rho$
meson mass. 
Other details of the simulation are described in
\cite{Aoki:2002uc}. 
We adopt the lattice NRQCD formalism
\cite{Thacker:1990bm,Lepage:1992tx} 
for heavy quarks. The non-perturbatively $O(a)$-improved
Wilson action \cite{Jansen:1998mx} is employed for both valence and
sea light quarks. 

We take five values of heavy quark mass $m_Q$ 
($am_Q$ = 1.3, 2.1, 3.0, 5.0 and 10.0) to cover 
$m_Q$= 3--20~GeV for the NRQCD action that contains all 
corrections of the order $1/m_Q$.
We take the valence light quark mass set equal to the sea quark
mass and then extrapolate to the physical $u$- and $d$-quark
masses, unlike the `partially quenched analysis' often adopted
in the literature.
The strange quark is treated in the quenched approximation.
There is an uncertainty in the determination of the
strange quark mass depending upon which strange hadron
is used as input.
We take three values for the hopping parameter $K_s$
= 0.13465, 0.13468 and 0.13491 and interpolate to 
$K_s(K)$ = 0.13486(3) (for the $K$ meson as input) and 
$K_s(\phi)$ = 0.13471(4) (for the $\phi$ meson as input)
\cite{Aoki:2002uc}.
The method to calculate $f_B$  and
the $B$ parameter follows our previous studies in the
quenched approximation \cite{Ishikawa:1999xu,Aoki:2002bh}.

\begin{figure}[tbp]
  \centering
  \includegraphics*[width=8cm,clip=true]{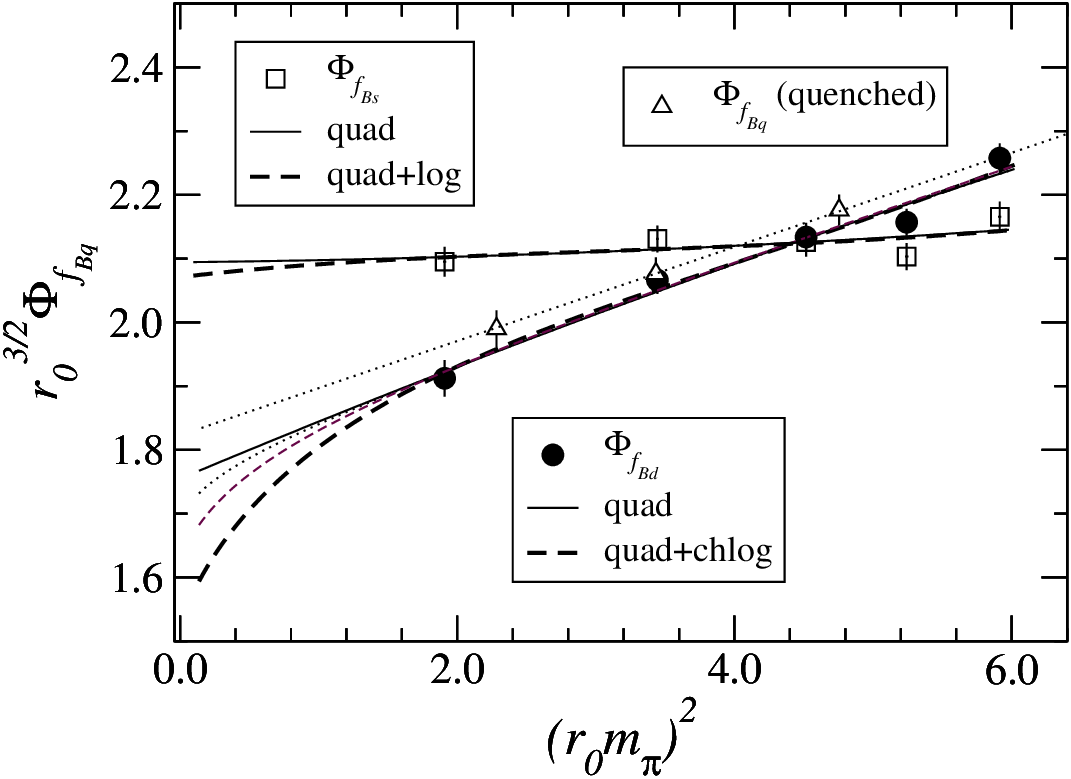}
  \caption{
    Chiral extrapolation of $\Phi_{f_{B_d}}$ (filled circles)
    and $\Phi_{f_{B_s}}$ (open squares).
    The quadratic extrapolation is shown by solid lines,
    while the fits with the hard cutoff chiral logarithm are
    shown for $\mu$ = 300 (dotted curve), 500 (thin dashed
    curve) and $\infty$ (thick dashed curve) MeV.
    Quenched results are also shown (triangles).
  }
  \label{fig:chi_fB}
\end{figure}

Figure~\ref{fig:chi_fB} shows the chiral extrapolation of
the decay constants $f_{B_q}$ expressed in terms of 
$\Phi_{f_{B_q}}\equiv f_{B_q}\sqrt{M_{B_q}}$
as a function of the pion mass squared.
For $f_{B_s}$, the result is shown at $K_s$=0.13465. 
In order to absorb the change of effective lattice scale 
that varies with $K_{sea}$ at a fixed bare coupling $\beta$,
both axes are normalized with $r_0$ determined from the
heavy quark potential at each sea quark mass. 
The heavy quark mass is interpolated to the $b$
quark using the lattice data.

Open triangles show quenched results obtained
at a similar lattice spacing $1/a$ = 1.83(2)~GeV 
($\beta$ = 6.0)
with the non-perturbatively improved Wilson quark action. 
Our observation that they lie close to the unquenched data
(filled circles) implies that the $B$ meson decay constant
takes a similar value in quenched and two-flavor QCD if the
scale is normalized by $r_0$. 
With the more conventional normalization of using the $\rho$ 
mass, however, the unquenched values are higher by about
20\% (see, e.g., \cite{Yamada:2002wh}), as is seen by
the fact that $r_0 m_\rho$ = 1.91(2) for $N_f$ = 2 while it is
2.20(3) on the quenched lattice \cite{Aoki:2002uc}.
This is understood as systematic errors of the quenched
approximation, with which the determination of the lattice scale
depends on which physical quantity is the input. With the
dynamical quarks these errors are significantly reduced,
leading to a convergent determination of the lattice scale.

The solid line represents a linear plus quadratic fit in
$(r_0m_\pi)^2$, which describes the lattice data well.
This fit, however, does not contain the chiral logarithmic term  
which is predicted by ChPT \cite{Grinstein:1992qt}:
\begin{eqnarray}
  \label{eq:chiral log}
  \frac{\Phi_{f_{B_d}}}{\Phi_{f_{B_d}}^{(0)}}
  = 1 
  - \frac{3(1+3g^2)}{4}\frac{m_\pi^2}{(4\pi f)^2}
  \ln\frac{m_\pi^2}{\mu^2}
  + \cdots,
\end{eqnarray}
where terms regular in $m_\pi^2$ are omitted, and
the coupling $g$ is the $B^*B\pi$ interaction in
ChPT. Lattice calculations \cite{deDivitiis:1998kj,Abada:2002xe}
give a value consistent with the empirical one measured for
$D^*\rightarrow D\pi$ decay,
$g = 0.59 \pm 0.01 \pm 0.07$ \cite{Anastassov:2002cw}.
Although there is an uncertainty in translating the value at
the $D^{(*)}$ meson to that at the infinitely heavy quark mass (where
the heavy-light ChPT is formulated), we take
$g$ = 0.6 to estimate the effect of the chiral
logarithm.

Let us here consider a simpler case.
For the pion decay constant the one-loop logarithmic term
is controlled by the number of dynamically active quark
flavors $N_f$ as 
$-\frac{N_f}{2} \frac{m_\pi^2}{(4\pi f)^2} \ln\frac{m_\pi^2}{\mu^2}$;
no uncertain parameters such as $g$ are involved.
Thus, the test for the presence of the chiral logarithm
is less ambiguous 
\cite{Hashimoto:2002vi}.
Our high statistics unquenched data, shown  
in Figure~2, 
exhibit quite a linear behavior for $(r_0m_\pi)^2>2$,
\textit{i.e.}, $m_\pi >$ 500~MeV; 
no appreciable curvature characteristic of
the chiral logarithm is observed.

\begin{figure}[tbp]
  \centering
  \includegraphics*[width=7.6cm,clip=true]{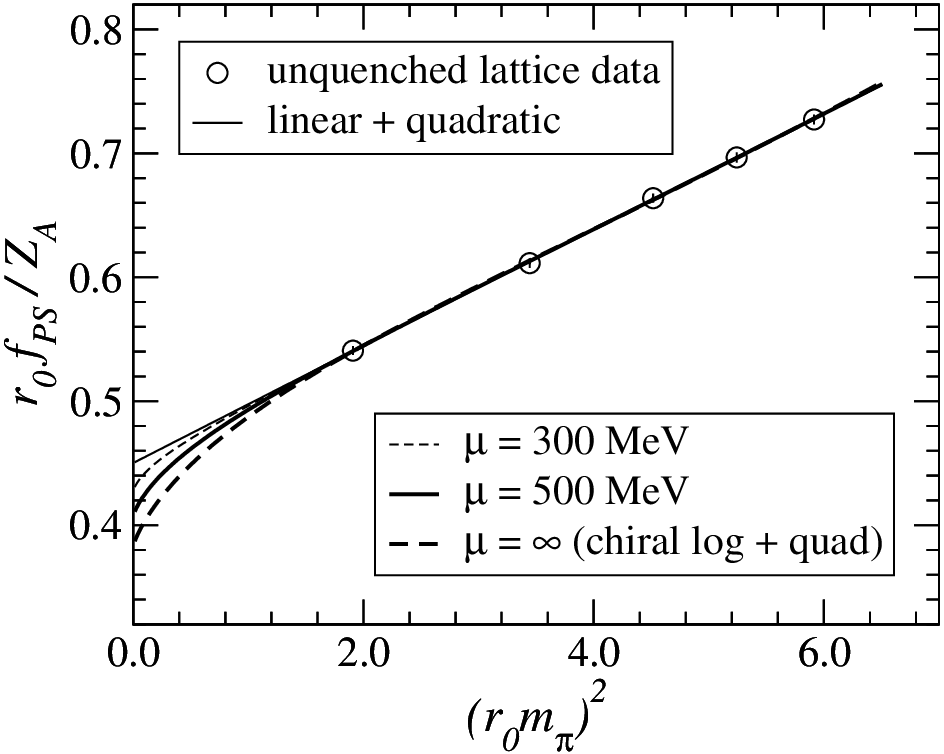}
  \caption{
    Chiral extrapolation of $f_\pi$ divided by the
    renormalization factor $Z_A$.
    The fits with the hard cutoff chiral logarithm are shown
    for $\mu$ = 300 (thin dashed curve), 500 (thick curve)
    and $\infty$ (dashed curve) MeV.
  }
  \label{fig:fpi}
\end{figure}

One may suspect that pions in the simulation are
too heavy to validate the use of ChPT.  
Another possibility may be the effect of explicit chiral symmetry
breaking of the Wilson quark action at finite lattice spacings, 
as was discussed recently in the context of ChPT 
by \cite{Aoki:2003yv} 
(see also \cite{Rupak:2002sm,Baer:2003mh}). Here we
explore the more naturally looking, former possibility that the
pion loop is suppressed for heavy pions and that the chiral
logarithm manifests itself only for sufficiently small sea quark
masses.
The authors of \cite{Detmold:2001jb} proposed a model that
incorporate such a behavior by introducing a hard cutoff 
regularization of the one-loop ChPT calculation. 
This amounts to the replacement,  
$m_\pi^2\ln m_\pi^2/\mu^2 \rightarrow
 m_\pi^2\ln m_\pi^2/(m_\pi^2+\mu^2)$,
where $\mu$ is the scale of the hard cutoff, beyond which
pion loop effects are suppressed.
This function has to be understood as a model when used
above the cutoff $\mu$.
We may use this model to explore the possible range of uncertainties
consistent with the lack of curvature in our data.  

Curves in Figure~\ref{fig:fpi} illustrate the chiral
extrapolation using the cutoff-logarithm plus a quadratic
term. 
All curves are consistent with the lattice data,
and they deviate from the polynomial only in the
small mass region.
The $\mu=\infty$ limit corresponds to the usual chiral
logarithm function plus a quadratic term, for which the
curvature cancels between the logarithmic and quadratic terms
in the region of simulations while giving a large effect below
$(r_0m_\pi)^2 <$ 1.
The other limit $\mu$ = 0~MeV corresponding to the
polynomial fit.
The variation with the parameter $\mu$
is taken as uncertainties in the chiral
extrapolation within this model. 
This shows that the value obtained with the polynomial fit
($\mu=0$~MeV), 
147(3)~MeV (here the errors are statistical only),
may be affected by a `hidden' chiral logarithm, leading to
128(2)~MeV if the effect is maximal ($\mu=\infty$~MeV).

A similar analysis can be made for the heavy-light decay
constant we have discussed above. The fits are shown in
Figure~\ref{fig:chi_fB} for $\mu$ = 300 and 500~MeV (thin
dotted curves) as well as for $\infty$~MeV (dashed curve).
The effect of the chiral logarithm can be $-$11\% for
$f_B$, if we take $\mu=\infty$ as an extreme case.
While this case is probably unrealistic, as it implies the 
validity of ChPT at very large mass scales,
we take this as a conservative estimate of the systematic
error, giving the lower limit for $f_B$. 
Other functional forms may also be adopted
(see, \textit{e.g.}, \cite{Sanz-Cillero:2003fq}),
but such models are expected to give numerically similar 
results in so far as the model is constrained by lattice data in 
the heavy pion mass region and by ChPT
in the light pion mass region. 

The effect of the chiral logarithm is small for
$f_{B_s}$, since the particle circulating the loop is kaon
or eta.
The formula in the partially quenched QCD is given
in \cite{Sharpe:1996qp}.
The chiral extrapolation is shown in Figure~\ref{fig:chi_fB}
with the lines for two extreme cases $\mu$ = 0 and
$\infty$~MeV.
The difference between the two is only 1\%.

To quote our results we take the central value from the
polynomial fit and include the variation in the presence of
the chiral logarithm as an error.
We obtain
\begin{eqnarray}
  f_{B_d} &= & 191(10)(^{+\ 0}_{-19})(12)(-)~ \mbox{MeV},\\
  f_{B_s} &= & 215(9)(^{+ 0}_{- 2})(13)(^{+6}_{-0})~ \mbox{MeV},\\
  \frac{f_{B_s}}{f_{B_d}} &=& 1.13(3)(^{+12}_{-\ 0})(2)(^{+3}_{-0}),
\end{eqnarray}
where the first error is statistical, 
the second is the uncertainty from the chiral extrapolation
and the other two are systematic errors explained in what
follows.
The error from the chiral extrapolation is one sided, since
the polynomial fit is taken as our central value. 
The systematic error given in the third parenthesis 
is those arising from the finite lattice spacing
(truncation of the actions and currents, and their
perturbative matching) and the truncation of terms higher
order in $1/m_Q$ in the NRQCD action. 
An order estimate of the truncation errors shows that the
most important contributions are
$O(\Lambda_{\mathrm{QCD}}^2/m_b^2)\sim$ 4\% and
$O(\alpha_s^2)\sim$ 4\%.
We add these errors by quadratures together with other
possible errors. 
In \cite{Ishikawa:1999xu}, it is shown for the quenched
lattice that such estimates correctly describe the error of
finite lattice spacing 
(see also \cite{Aoki:2002bh} for $B_B$).
The errors we obtained is consistent with those in the
quenched case at the comparable lattice spacing. 
The errors in the last parenthesis represent the ambiguity
in the determination of the strange quark mass, for which we
adopt the value with the $K$ mass as the central value.

For the $B$ parameter, ChPT predicts $-(1-3g^2)/2$ for the
coefficient of the chiral log term instead of 
$3(1+3g^2)/4$ in (\ref{eq:chiral log}) \cite{Sharpe:1996qp}.
Therefore, the effect of the chiral logarithm is 
negligible in practice.
For $B_{B_s}$ there is no chiral logarithm as a function of
sea quark mass.

\begin{figure}[tbp]
  \centering
  \includegraphics*[width=8cm,clip=true]{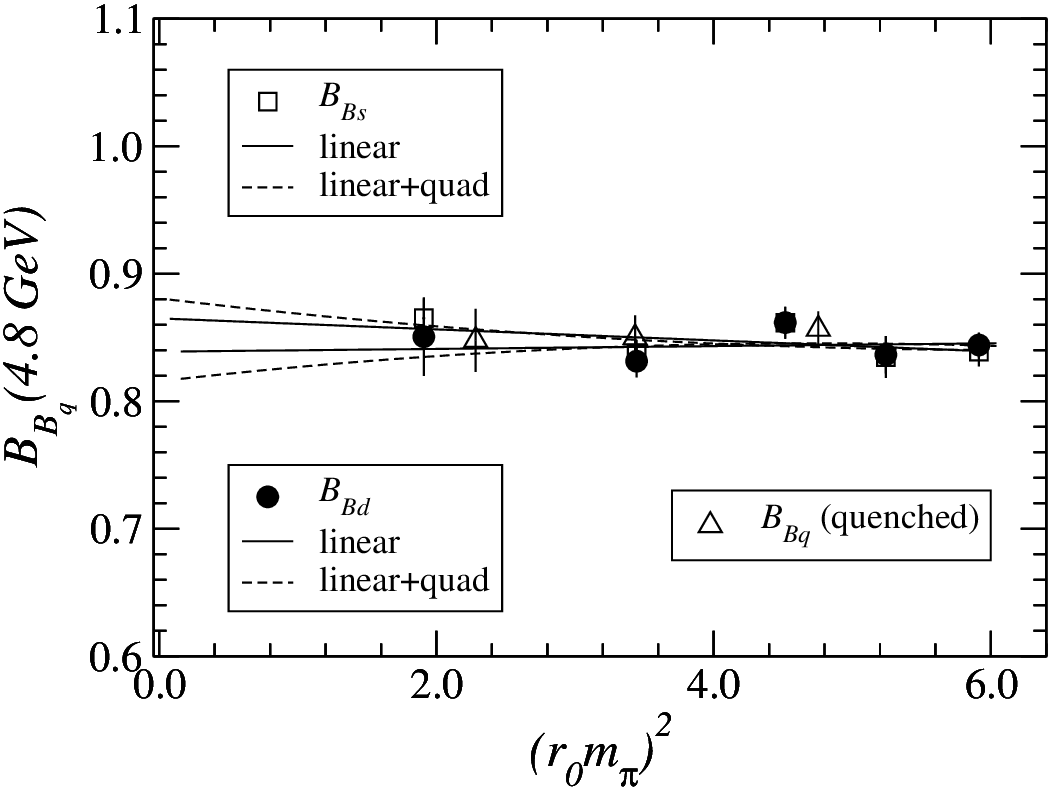}
  \caption{
    Chiral extrapolation of $B_{B_d}$ (filled circle)
    and $B_{B_s}$ (open square).
    The four-quark operator is perturbatively renormalized,
    and the ``method I'' is employed for the truncation of
    higher order corrections (in $\alpha_s$ and $1/m_b$) as
    an example \cite{Aoki:2002bh}.
    }
  \label{fig:chi_BB}
\end{figure}

Figure~\ref{fig:chi_BB} shows the chiral extrapolation of
$B_{B_q}(\mu_b)$ at $\mu_b$ = $m_b$ (= 4.8~GeV) and the fits
without the chiral logarithm. 
The triangles show the quenched results.
The sea quark effect is small for this quantity.

Our unquenched results obtained with a linear chiral
extrapolation are
\begin{eqnarray}
  B_{B_d}(m_b) &=& 0.836(27)(^{+\ 0}_{-27})(56)(-),\\
  B_{B_s}(m_b) &=& 0.850(22)(^{+18}_{-\ 0})(57)(^{+5}_{-0}),\\
  \frac{B_{B_s}}{B_{B_d}}
               &=& 1.017(16)(^{+53}_{-\ 0})(17)(^{+6}_{-0}).
\end{eqnarray}
The meaning of errors is the same as for $f_B$, except for
the second one, 
\textit{i.e.}, those associated with the chiral
extrapolation:
we take the central value from the linear fit and put the
difference from the linear plus quadratic fit as the
systematic error. 

The amplitude of neutral $B$ meson mixing is
proportional to $f_B^2 B_B$.
Using the conventionally adopted renormalization-scale
independent definition $\hat{B}_B$, which is related to $B_B(m_b)$ as
$\hat{B}_B = 1.528 B_B(m_b)$
for $\Lambda^{(5)}_{\overline{\rm MS}}$ = 225~MeV,
we find
\begin{eqnarray}
  \label{eq:fB_sqrtBB}
  f_{B_d}\sqrt{\hat{B}_{B_d}} 
  &=& 215(11)(^{+\ 0}_{-23})(15)(-)~ \mbox{MeV},\\
  f_{B_s}\sqrt{\hat{B}_{B_s}}
  &=& 245(10)(^{+3}_{-2})(17)(^{+7}_{-0})~ \mbox{MeV},
\end{eqnarray}
and for the SU(3) breaking ratio $\xi$,
\begin{equation}
  \label{eq:xi}
  \xi = 1.14(3)(^{+13}_{-\ 0})(2)(^{+3}_{-0}).
\end{equation}
The chiral extrapolation gives the largest entry of
systematic errors for $\xi$, as also suggested in
\cite{Kronfeld:2002ab}.
Compared to the commonly assumed number in the
phenomenological analysis, $f_{B_d}\sqrt{\hat{B}_{B_d}}$ =
230(28)(28)~MeV (\textit{e.g.}, \cite{Hocker:2001xe}), where 
the second error is the quenching uncertainty, 
our central value of (\ref{eq:fB_sqrtBB}) is slightly lower and the 
quenching error is eliminated.

In conclusion we have obtained unquenched lattice estimate
of the $B^0-\bar{B}^0$ mixing matrix elements, including the
decay constant, in a consistent set of simulations.  
Although the simulation is made at a relatively large mass of
dynamical quarks, we explored the range of errors associated
with the chiral extrapolation:
we expect that the true values of the matrix elements are within
the range of indicated errors, even if the chiral logarithm 
would become manifest at a small quark mass.

\begin{acknowledgments}
This work is supported by Large Scale Simulation Program
No.~79 (FY2002) of High Energy Accelerator Research
Organization (KEK), 
and also 
in part by the Grants-in-Aid of the Ministry of Education 
(Nos.
12740133,
13135204,
13640259,
13640260,
14046202,
14540289,
14740173,
15204015,
15540251,
15540279).

\end{acknowledgments}


\end{document}